\documentclass[aps,prl,twocolumn,preprintnumbers,superscriptaddress,amsmath]{revtex4}

\usepackage{amssymb}
\usepackage{txfonts}
\usepackage{amssymb}
\usepackage{graphicx}
\usepackage{subfigure}
\usepackage{dcolumn}
\usepackage{bm}
\usepackage{color}
\usepackage{upgreek}

\begin{document}

\title{Chip-to-chip quantum photonic controlled-NOT gate teleportation}

\author{Lan-Tian Feng}
\affiliation
{CAS Key Laboratory of Quantum Information, University of Science and Technology of China, Hefei 230026, China.}
\affiliation{CAS Synergetic Innovation Center of Quantum Information $\&$ Quantum Physics, University of Science and Technology of China, Hefei 230026, China.}
\author{Ming Zhang}
\affiliation{State Key Laboratory for Modern Optical Instrumentation, Centre for Optical and Electromagnetic Research,
Zhejiang Provincial Key Laboratory for Sensing Technologies, Zhejiang University, Zijingang Campus, Hangzhou 310058, China.}
\author{Di Liu}
\author{Yu-Jie Cheng}
\author{Xin-Yu Song}
\affiliation
{CAS Key Laboratory of Quantum Information, University of Science and Technology of China, Hefei 230026, China.}
\affiliation{CAS Synergetic Innovation Center of Quantum Information $\&$ Quantum Physics, University of Science and Technology of China, Hefei 230026, China.}
\author{Yu-Yang Ding}
\affiliation{Hefei Guizhen Chip Technologies Co., Ltd., Hefei 230000, China.}
\author{Dao-Xin Dai}
\affiliation{State Key Laboratory for Modern Optical Instrumentation, Centre for Optical and Electromagnetic Research,
Zhejiang Provincial Key Laboratory for Sensing Technologies, Zhejiang University, Zijingang Campus, Hangzhou
310058, China.}
\author{Guo-Ping Guo}
\author{Guang-Can Guo}
\author{Xi-Feng Ren\footnote[2]{renxf@ustc.edu.cn}}
\affiliation
{CAS Key Laboratory of Quantum Information, University of Science and Technology of China, Hefei 230026, China.}
\affiliation{CAS Synergetic Innovation Center of Quantum Information $\&$ Quantum Physics, University of Science and Technology of China, Hefei 230026, China.}

\begin{abstract}
Quantum networks provide a novel framework for quantum information processing, significantly enhancing system capacity through the interconnection of modular quantum nodes.
Beyond the capability to distribute quantum states, the ability to remotely control quantum gates is a pivotal step for quantum networks.
In this Letter, we implement high fidelity quantum controlled-NOT (CNOT) gate teleportation with state-of-the-art silicon photonic integrated circuits. 
%
Based on on-chip generation of path-entangled quantum state, CNOT gate operation and chip-to-chip quantum photonic interconnect, the CNOT gate is teleported between two remote quantum nodes connected by the single-mode optical fiber. 
Equip with 5$\,$m (1$\,$km)-long interconnecting fiber, quantum gate teleportation is verified by entangling remote qubits with 95.69\%$\,\pm\,$1.19\% (94.07\%$\,\pm\,$1.54\%) average fidelity and gate tomography with 94.81\%$\,\pm\,$0.81\% (93.04\%$\,\pm\,$1.09\%) fidelity. 
%
%
These results advance the realization of large-scale and practical quantum networks with photonic integrated circuits.


\end{abstract}
\maketitle

In quantum networks, quantum information is transmitted, processed and stored among different quantum nodes, which overcome the limitation of individual physical setups and provide a novel framework to scale up the system capacity. 
Except for quantum communication, quantum networks also play a crucial role in quantum computing and quantum metrology.
In the distributed manner, numerous quantum applications such as blind quantum computing \cite{Broadbent2009,Barz2012}, quantum-enhanced long-baseline interferometry \cite{Rajagopal2024} and atomic clock synchronization \cite{Kómárl2014} have emerged.
In order to implement quantum networks that comprise a substantial number of quantum nodes, it is imperative that each individual node features the scalable and integrated quantum hardware.
Furthermore, high-capacity quantum interconnects should be established among these nodes.

%
%

Quantum photonic integrated circuits (QPICs) manipulate photons in phase-stable circuitry with a millimeter-scale footprint \cite{Wang2020,Feng2022}. 
QPICs will provide an integrated and scalable architecture to implement large-scale and practical quantum networks. 
In these networks, photons link multiple quantum nodes via long-distance and complex-medium quantum channels.
As an example, integrated quantum photonic sources, recognized for their high brightness and broad bandwidth \cite{Feng2020}, play an important role in combination with wavelength division multiplexing/demultiplexing systems for state distribution and multiuser quantum networks \cite{Liu2022,Wen2022,Ren2023}.
There are varieties of on-chip structures such as two-dimensional grating coupler \cite{Wang2016}, grating coupler array \cite{Ding2017}, and polarization beam rotator and combiner (PBRC) \cite{Wei2020} have been developed for quantum state fan-in/fan-out of the chip. 
Multicore fiber \cite{Ding2017} and few-mode fiber \cite{Lu2024} have been used to achieve high-dimensional quantum photonic interconnect. 
With more fundamental building blocks are further assembled and cascaded on one single chip, chip-to-chip quantum key distribution \cite{Wei2020,Ding2017,Sibson2017,Paraïso2021}, entangled quantum state distribution \cite{Wang2016,Alexander2024} and teleportation \cite{Llewellyn2020} and multichip multidimensional quantum entanglement network \cite{Zheng2023} have been experimentally demonstrated.

To further innovate quantum networks with QPICs, it is crucial to perform quantum gate operation between qubits in separate quantum nodes. 
A cornerstone task in this challenge is implementing high fidelity teleportation of controlled-NOT (CNOT) gate. 
The CNOT gate teleportation operation entangles remote qubits and is an essential tool for universal quantum computation \cite{Gottesman1999}.
Achieving the CNOT gate teleportation requires sharing an entangled photon-pair between two quantum nodes and needs complex on-chip linear optical quantum manipulation within each node \cite{Gasparoni2004,Huang2004,Gao2010}, which present significant challenges to the device performance.

\begin{figure*}[t]
\centering
\includegraphics[width=14.0cm]{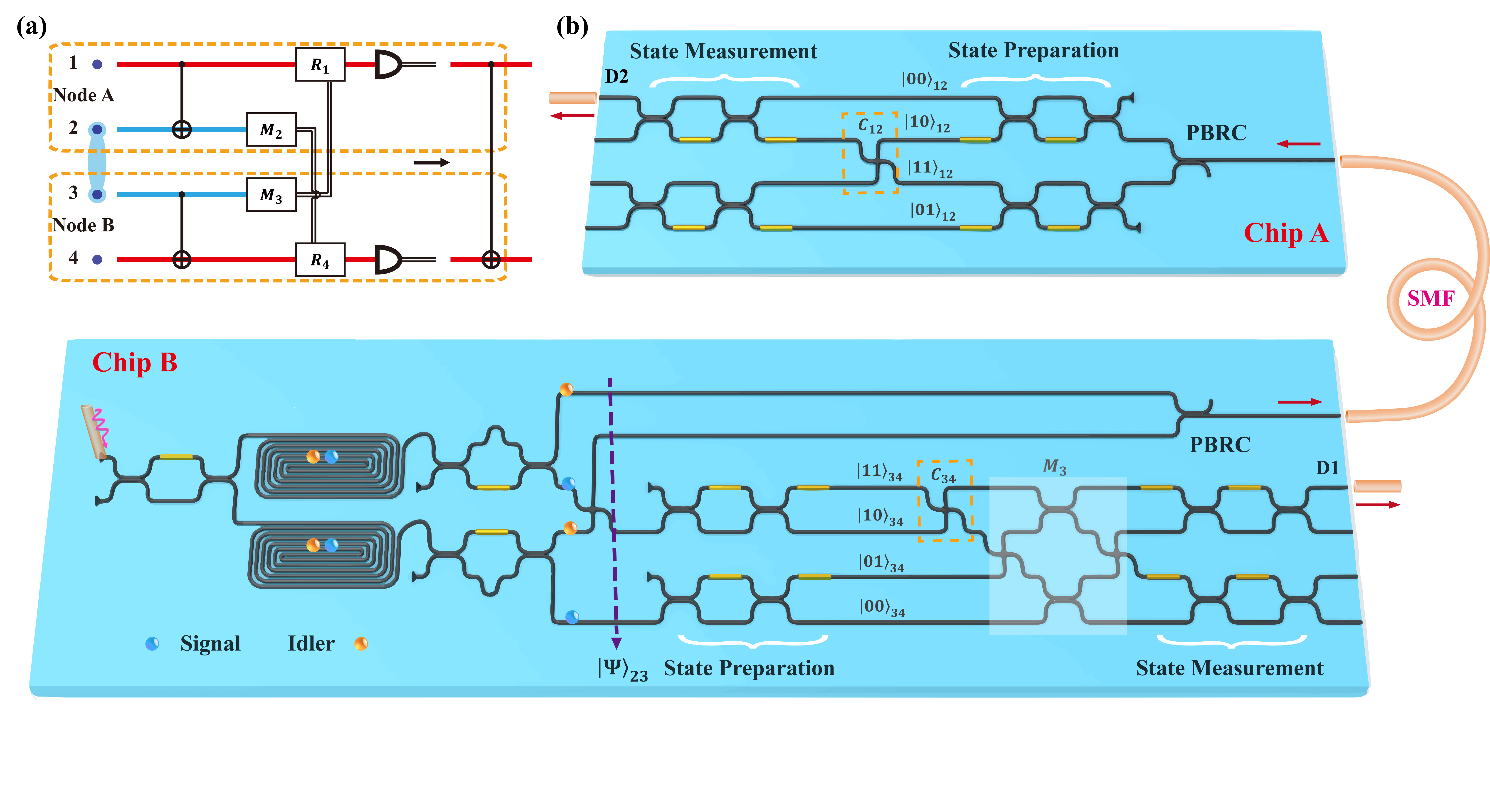}
\caption {The idea of quantum CNOT gate teleportation and schematic chip-to-chip realization. 
(a) A sketch to show the basic idea. 
$M_2$ and $M_3$ represent single qubit measurements on corresponding qubits, and $R_1$ and $R_4$ represent corresponding single-qubit rotations based on the measurement results. 
(b) Schematic diagram of the chip-to-chip quantum CNOT gate teleportation. 
Chip A includes qubits 1\&2, and chip B includes qubits 3\&4.
Chips A and B share an entangled Bell pair (qubits 2\&3) and constitute a two-node quantum network.
Photons at ports D1\&D2 are collected and detected for two-photon coincidence measurement.
PBRC: polarization beam rotator and combiner. SMF: single-mode optical fiber.}
\label{fig1}
\end{figure*}

Here, we report the chip-to-chip quantum photonic CNOT gate teleportation using state-of-the-art silicon PICs.
These silicon PICs demonstrate compatibility with CMOS fabrication and pave the way for developing large-scale and modular quantum photonic devices \cite{Harris2016,Rudolph2017,Alexander2024}. 
We prepare entangled quantum photonic source on a silicon chip and transmit one of the photons to another chip via a single-mode optical fiber link.
Through linear optical quantum manipulation in each node, a quantum CNOT gate is teleported from acting on local qubits to acting on remote qubits.
We confirm the quality of the teleportation process by entangling path-encoded remote qubits and quantum process tomography.
With 5$\,$m-long interconnecting fiber, an average entangled state fidelity of 95.69\%$\,\pm\,$1.19\% and a process fidelity of 94.81\%$\,\pm\,$0.81\% are achieved. 
Additionally, we extend the optical fiber length to 1$\,$km. 
Even without employing any feedback techniques, the setup maintains remarkable stability. 
%
The obtained average entangled state fidelity is 94.07\%$\,\pm\,$1.54\% and process fidelity is 93.04\%$\,\pm\,$1.09\%. 
The chip-to-chip quantum CNOT gate teleportation we demonstrated here can be adapted to more complex scenarios, such as increased distances, a greater number of quantum nodes, and field environment, and thereby will play an important role in practical quantum networks. 

The basic idea of the CNOT gate teleportation is shown in Fig. 1(a), which has minimum resource cost and requires only local operation, classical communication, and a single entangled pair shared between two quantum nodes \cite{Eisert2000}.
In this simple network, each node includes two qubits, and qubits 2\&3 are entangled.
The action of a CNOT gate performed on local qubits 1\&2 ($C_{12}$) can be teleported to operate equivalently on remote qubits 1\&4 ($C_{14}$).
To achieve this goal, extra local CNOT gate $C_{34}$ should be executed on qubits 3\&4, along with single qubit measurements on qubits 2\&3 ($M_{2}$ and $M_{3}$) and rotations on qubits 1\&4 ($R_{1}$ and $R_{4}$) based on the measurement results. 
Assume the entangled pair has the form of
$
|\Psi\rangle_{23}=(\left|00\right\rangle_{23}+\left|11\right\rangle_{23})/\sqrt{2},
$
the evolution of the quantum state follows the identity:
\begin{equation}
C_{34}C_{12}|\Phi\rangle_{14}|\Psi\rangle_{23}=\sum_{i,j}|i\rangle_{2}|j\rangle_{3}R_{1}(i,j)R_{4}(i,j)C_{14}|\Phi\rangle_{14},
\end{equation} 
where $|\Phi\rangle_{14}=|\phi\rangle_{1}\otimes|\phi\rangle_{4}$, $|i\rangle_{2}=|0\rangle_{2}, |1\rangle_{2}$ and $|j\rangle_{3}=|\pm\rangle_{3}=(\left|0\right\rangle_{3}\pm\left|1\right\rangle_{3})/\sqrt{2}$.
More details about the CNOT gate teleportation is given in Supplemental Materials.


The schematic diagram of CNOT gate teleportation between two chip-based quantum nodes is shown in Fig. 1(b).
The integrated photonic node comprises a range of fundamental components, including optical grating and edge coupler, multimode interference (MMI) beamsplitter, waveguide crosser, thermal-optic phase shifter (PS) and PBRC.
MMI beamsplitters and PSs form balanced Mach-Zehnder interferometers (MZIs) for quantum state manipulation and unbalanced MZIs (UMZIs) for separating signal and idler photons.
The PBRC is utilized for path-polarization interconversion, which merges two TE input optical signals into a single output port and converts one signal to TM mode, or vice versa.


In the experiment, we use one continues wave laser centered at 1550.11$\,$nm as the pump, and the pump power of the laser is kept at 8$\,$mW (before coupling to the chip).
The laser is coupled into the chip (Chip B) via a grating coupler, and one MZI is used to split the laser equally into two 2.5$\,$cm-long single-mode spiral silicon waveguides.
%
%
In the waveguides, two pump photons are annihilated, and signal–idler photon pairs are generated with the spontaneous four-wave mixing process \cite{Feng2019}.
After passing through the spiral silicon waveguide, signal and idler photons in both paths are separated by using two UMZI filters. 
Based on the filtering property of UMZI (see Supplemental Materials), we select signal and idler photons with respective central wavelengths of 1538.18$\,$nm and 1562.23$\,$nm, and with 200$\,$GHz bandwidth.
At the outputs of the filters, one waveguide crosser swaps the middle two paths where signal and idler photons are located. 
By this way, we obtain the path-entangled photon-pair source to encode qubits 2\&3.
%
To establish an path-entangled quantum network between two integrated photonic chips, we utilize one PBRC to convert the encoding method of idler photons from path to polarization.
Then, idler photons with polarization encoding are coupled into single-mode optical fiber (5$\,$m length) via the edge coupler, and are delivered into another chip (Chip A).
On the receiving chip, the encoding method of idler photons is restored to path by another PBRC.

In order to encode two qubits by using the single photon in each quantum node, the path dimension of each photon in the entangled photon-pair source is expanded to four, and the quantum states represented by each path are also indicated in Fig. 1(b). 
The desired quantum state of qubits 1\&4 $|\Phi\rangle_{14}$ can be achieved through adjusting PSs located within the designated 'State Preparation' area.
%
Local CNOT gates $C_{12}$ and $C_{34}$ are realized with the same procedure, and a waveguide crosser in the path $|1\rangle$ of the control qubit is used to swap the path state of the target qubit. 
%
The measurement of qubit 2 in the basis $\{|0\rangle, |1\rangle\}$ is easily done by detecting output ports of the chip with a single-photon detector.
The measurement of qubit 3 in the basis $\{|+\rangle, |-\rangle\}$ requires that the photon paths $|0\rangle_3$ and $|1\rangle_3$ interfere at the MMI beamsplitter before the single-photon detection, and we employ a network comprising of MMI beamsplitters and waveguide crossers ($M_3$ shown in Fig. 1(b)) to accomplish this objective. 
After performing the basis conversion process, collecting photons from a combination of various output ports on two separate chips enables the detection of the states $|0+\rangle_{23}$, $|0-\rangle_{23}$, $|1+\rangle_{23}$ and $|1-\rangle_{23}$, respectively. 
For simplicity, we just collect photons in the term $|0+\rangle_{23}$ (at ports D1, D2 in Fig. 1(b)) for following coincidence measurement and corresponding single-qubit rotations on qubits 1\&4 are $R_1=R_4=I$.
Before each single-photon detection, MZI and PSs are positioned in the designated 'State Measurement' area to analyze the final state $|\Phi\rangle_{14}$.


%


\begin{figure}[t]
\centering
\includegraphics[width=8.0cm]{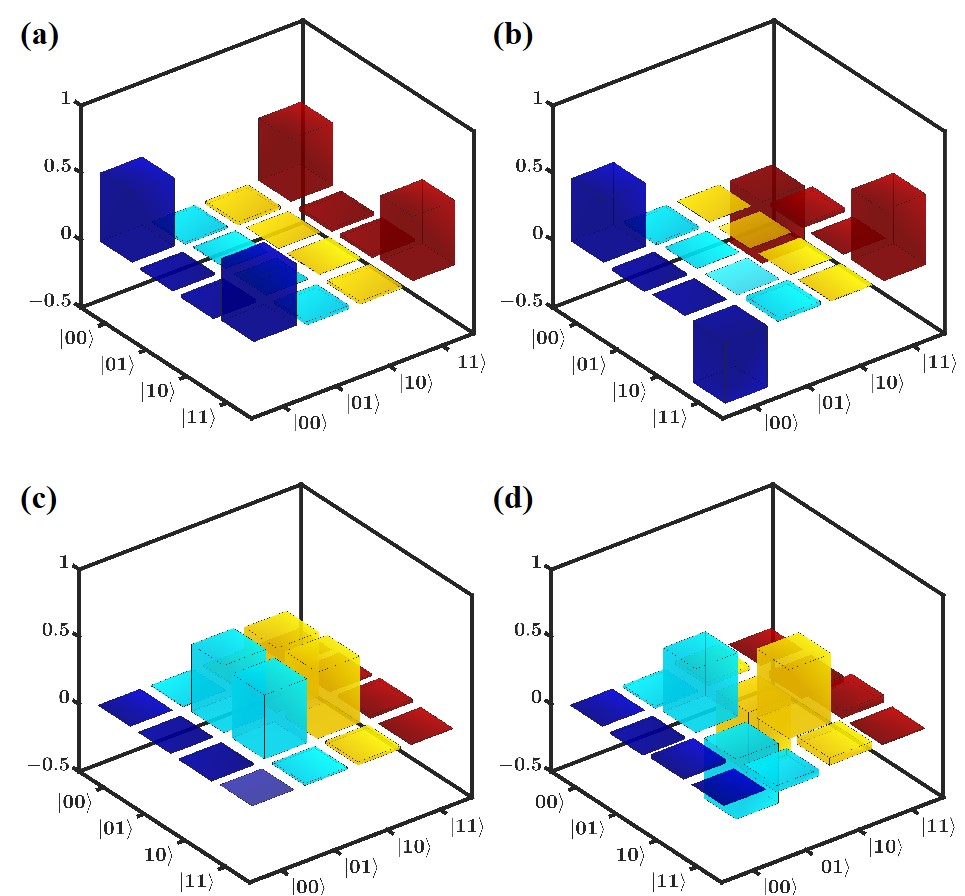}
\caption {\textbf{} Quantum state tomography for the generated Bell states between qubits 1\&4. Real parts of the measured density matrices of output states, with different input states of \textbf{a.} $\left|+\right\rangle_1\left|0\right\rangle_4$, \textbf{b.} $\left|-\right\rangle_1\left|0\right\rangle_4$, \textbf{c.} $\left|+\right\rangle_1\left|1\right\rangle_4$, and \textbf{d.} $\left|-\right\rangle_1\left|1\right\rangle_4$. The imaginary parts of the measured density matrices are negligible. The average fidelity is obtained as $\bar{F}=95.69\%\,\pm1.19\%$, which confirms the entangling function of the teleported CNOT gate.}
\label{fig3}
\end{figure}

\begin{figure*}[t]
\centering
\includegraphics[width=13.0cm]{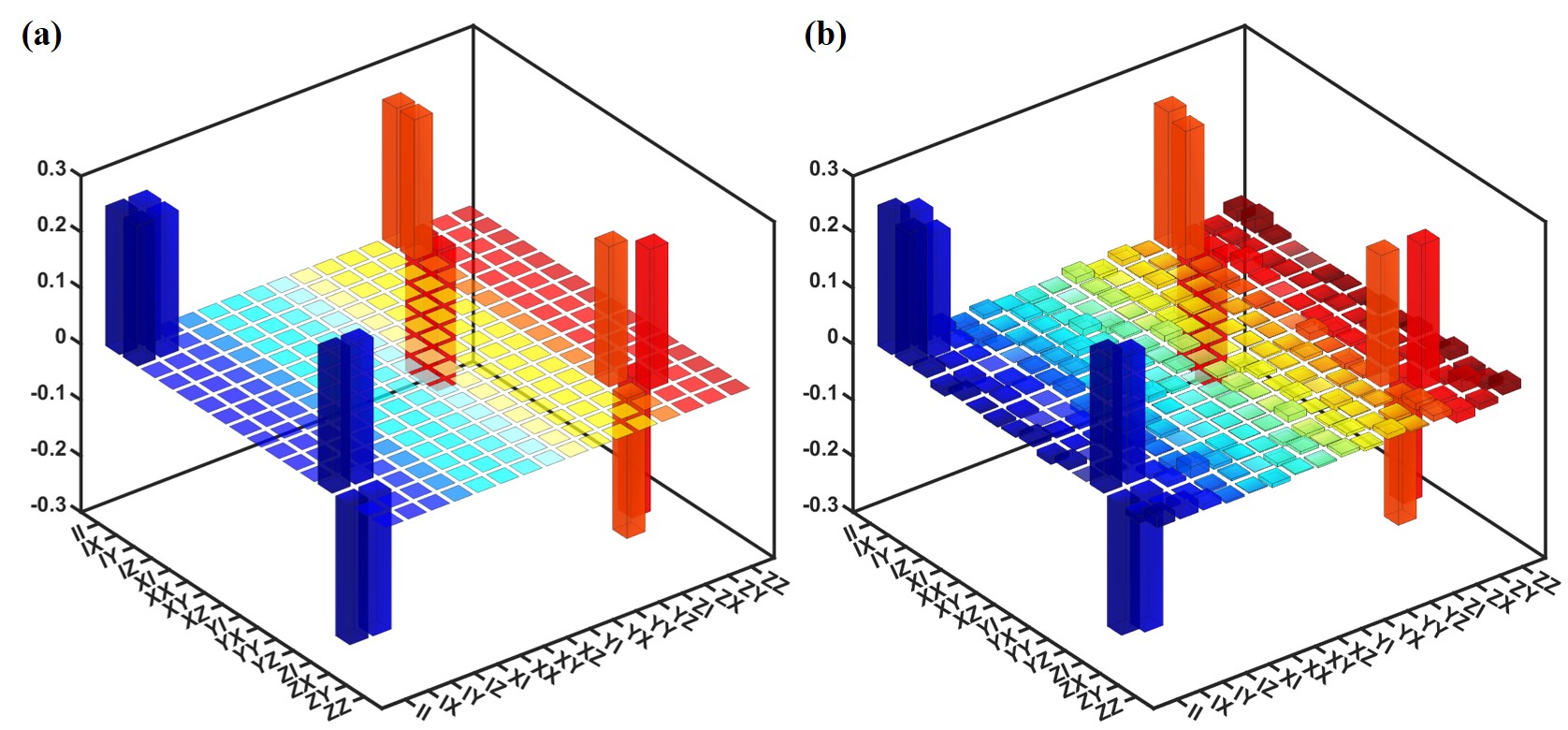}
\caption {\textbf{}Real part of the 16$\times$16 process matrix for the chip-to-chip teleported CNOT gate. (a) The ideal result $\chi_{\rm{ideal}}$. (b) The experimental result $\chi_{\rm{exp}}$ reconstructed from the raw data. 
Here, $\bm{I}$ represents the identity matrix, and $\bm{X}$, $\bm{Y}$, and $\bm{Z}$ represent the Pauli matrices $\sigma_{\rm{X}}$, $\sigma_{\rm{Y}}$, and $\sigma_{\rm{Z}}$, respectively.}
\label{fig4}
\end{figure*}

Throughout the experiment, one key step is to ensure high-fidelity chip-to-chip quantum photonic interconnect. 
%
We utilize a combination of PBRC and edge coupler to achieve path-polarization interconversion.
The PBRC is designed based on the ref. \cite{Sacher2014}, and its estimated losses for TE input to TM (TE) output or vise versa are lower than 0.4 (0.6)$\,$dB.
Besides, we use an ultra-high numerical aperture (UNHA) optical fiber to couple with the edge coupler, and the other end of the UHNA fiber is spliced to common single-mode fiber (SMF-28). 
The small mode field diameter of UHNA fiber matches well with the output mode field of the edge coupler, which facilitating efficient chip-fiber coupling.  
Specifically, TM (TE) losses of 3.42 (3.57)$\,$dB per coupler from chip to common optical fiber are achieved.

To maintain the quantum state within the interconnect fiber, we generate a sequence of bright light polarization states by programmatically configuring the interferometer preceding the spiral waveguides on Chip B. 
Upon transmission of these states to Chip A via the fiber, we proceed with state analysis on Chip A.
For each transmitted state, we evaluate both the state itself and its orthogonal counterpart.
Simultaneously, we continuously fine-tune one polarization controller situated in the fiber until the measurement outcomes align with the expected results of the transmitted state.
For the transmitted quantum state, the isolation degree is defined as the ratio of the maximum outcome to the minimum outcome.
After repeating above procedures multiple times, the isolation degree of each transmitted state exceeds 200 in the chip-to-chip photonic interconnect process.  

We demonstrate that the constructed quantum network exhibits exceptionally high quality by analyzing the chip-to-chip distributed entangled state $|\Psi\rangle_{23}$.
Photons on two separate chips are directed into paths labeled as $\left|00\right\rangle_{12}$, 
$\left|11\right\rangle_{12}$, $\left|00\right\rangle_{34}$ and $\left|10\right\rangle_{34}$ by configuring PSs. 
We first measure entanglement correlation fringes in the quantum network, and achieve average interference visibility of 95.49\%$\,\pm\,$1.17\%. 
This result verifies the presence of Bell nonlocality \cite{Rarity1990} and signifies that a reliable chip-based entangled quantum network has been successfully established. 
%
%
Then, we perform quantum state tomography \cite{James2001} to measure the state density matrix $\hat\rho$ between two quantum nodes.  
By configuring PSs, we achieve arbitrary single-qubit measurement capabilities, which allows us to perform all 16 measurements $\{\left|0\right\rangle,\left|1\right\rangle,\left|0\right\rangle+\left|1\right\rangle,\left|0\right\rangle+i\left|1\right\rangle\}^{\otimes 2}$ that required to reconstruct the density matrix of the state.
The fidelity between the ideal density matrix $\hat\rho_{\rm{ideal}}$ of the state $|\Psi\rangle_{23}$ and the measured density matrix $\hat\rho_{\rm{mea}}$ is defined as $F=\rm{Tr}(\hat\rho_{\rm{mea}}\hat\rho_{\rm{ideal}})$, where $\rm{Tr}$ represents the trace.
We obtain a raw fidelity of 95.76\%$\,\pm\,$0.83\%, confirming that the entangled quantum state is prepared with a high quality on Chip B and faithfully transferred to Chip A. 
%
More results about the quantum network verification are given in Supplemental Materials.

Furthermore, we characterize the performance of the teleportated CNOT gate in the quantum network.  
After preparing four logical basis states $\left|00\right\rangle_{14}$, 
$\left|01\right\rangle_{14}$, $\left|10\right\rangle_{14}$, $\left|11\right\rangle_{14}$ and measuring the probability
for each state, we obtain the truth table of the CNOT gate with a fidelity of 98.76\%$\,\pm\,$0.34\%. 
The basic function of the teleportated CNOT gate is to entangle remote qubits 1\&4. 
Here we show this function with our chip-based quantum network, with the input state of $\{\left|\pm\right\rangle_1\left|0\right\rangle_4, \left|\pm\right\rangle_1\left|1\right\rangle_4\}$. 
If the CNOT gate is perfect, input states are expected to be converted into four maximally entangled Bell states:
\begin{equation}
\begin{aligned}
|\Phi^{\pm}\rangle=\frac{1}{\sqrt{2}}(\left|00\right\rangle_{14}\pm\left|11\right\rangle_{14}),\\
|\Psi^{\pm}\rangle=\frac{1}{\sqrt{2}}(\left|01\right\rangle_{14}\pm\left|10\right\rangle_{14}).
\end{aligned}
\end{equation}
%
The output states are collected and reconstructed by performing quantum state tomography. 
We present the reconstructed density matrices $\hat\rho_{\rm{mea}}$ in Fig. \ref{fig3}, and calculate the raw fidelities for different output states as $F_{|\Phi^{+}\rangle}=95.86\%\,\pm\,1.12\%$, $F_{|\Phi^{-}\rangle}=95.52\%\,\pm1.01\%$, $F_{|\Psi^{+}\rangle}=97.30\%\,\pm1.81\%$ and $F_{|\Psi^{-}\rangle}=94.07\%\,\pm0.80\%$. 
The average fidelity is $\bar{F}=95.69\%\,\pm1.19\%$, which confirms the good entangling function of the teleported CNOT gate. 

To fully characterize the CNOT gate teleportation process, quantum process tomography is carried out. 
Any dynamic process can be seen as a positive map $\varepsilon$ from an arbitrary input state $\hat\rho_{\rm{in}}$ to output state $\hat\rho_{\rm{out}}$ \cite{Brien2004,White2007,Feng20222}, which can be written as:
\begin{equation}
\hat\rho_{\rm{out}}=\varepsilon(\hat\rho_{\rm{in}})=\sum_{\rm{m,n}=1}^{d^2}\chi_{\rm{mn}}\hat{A}_{\rm{m}}\hat\rho_{\rm{in}}\hat{A}^{\dagger}_{\rm{n}},
\end{equation} 
In the formula, $\hat{A}_{\rm{m(n)}}$ represent a series of operators acting on arbitrary input state $\hat\rho_{\rm{in}}$, with $d^2=16$ for two-qubit system, and symbol $\dagger$ represents conjugate transpose operation.
Thus, the matrix $\chi$ completely and uniquely characterizes $\varepsilon$ with respect to the basis $\hat{A}_{\rm{m,n}}$ and contains all the information of the dynamic process.
We prepare 16 different input states from the set $\{\left|0\right\rangle,\left|1\right\rangle,\left|0\right\rangle+\left|1\right\rangle,\left|0\right\rangle+i\left|1\right\rangle\}^{\otimes 2}$ for qubits 1\&4, and quantum state tomography is performed for each output state. 
In total, 256 projections are carried out to fully characterize the CNOT gate teleportation operation. 
The experimentally reconstructed process matrix $\chi_{\rm{exp}}$ is presented in Fig. 3, together with the ideal one $\chi_{\rm{ideal}}$. 
With the definition of $F=\rm{Tr}(\chi_{\rm{exp}}\chi_{\rm{ideal}})$, the process fidelity is obtained as $F=$94.81\%$\,\pm\,$0.81\%, which is comparable to the reported results for the photonic CNOT gate on one single chip \cite{Zhang2021,Santagati2017} and is sufficient for further chip-to-chip quantum information processing \cite{Gasparoni2004}.

At last, we extend length of the interconnecting optical fiber to 1$\,$km, and such distance is necessary for practical quantum networks.
Apart from the introduced insertion loss of 0.6 dB, the long optical fiber makes minimal impact on the teleported CNOT gate between two chip-based quantum nodes.
We also characterize the gate teleportation operation with quantum state tomography and process tomography, and achieve average entangled state fidelity of 94.07\%$\,\pm\,$1.54\% and process fidelity of 93.04\%$\,\pm\,$1.09\%. 
More detailed results about the quantum network with 1$\,$km-long optical fiber link are provided in Supplemental Materials.

Gate teleportation operation represents a critical step for quantum networks.
In this work, we establish two chip-based quantum nodes in silicon and demonstrate high fidelity chip-to-chip quantum photonic CNOT gate teleportation.
The single-mode optical fiber link exhibits high capacity and low loss while the stability and scalability of chip-based quantum nodes reduce the cost and complexity of quantum networks, thereby enhancing their flexibility and efficiency.
Further research can improve the chip-fiber coupling efficiency by optimizing the on-chip structure.
The large-scale and reconfigurable QPIC enables us to incorporate more qubits into each node, and to demonstrate gate teleportation operation among multiple nodes as well as more complex distributed quantum information processing.
The optical fiber link we utilize is capable of being extended in length and can be deployed into noisy field environments, making it suitable for practical quantum applications.
Moreover, the source, filter and heralding single-photon detector can be cointegrated in silicon QPIC \cite{Alexander2024}, which promises higher levels of integration.
In conclusion, we have prepared an on-chip path-entangled photon-pair source, sent one photon to another chip, and demonstrated chip-to-chip quantum CNOT gate teleportation.
The entire procedure exhibits high fidelity and is poised for further quantum information applications with integrated quantum nodes.
Therefore, this work advances the realization of large-scale and practical quantum networks.

This work was supported by National Key Research and Development Program of China (2022YFA1204704), the National Natural Science Foundation of China (NSFC) (62061160487, 62275240, T2325022, U23A2074, 62321166651), the CAS Project for Young Scientists in Basic Research (No. YSBR-049), Key Research and Development Program of Anhui Province (2022b1302007) and the Fundamental Research Funds for the Central Universities. This work was partially carried out at the USTC Center for Micro and Nanoscale Research and Fabrication.


\end{document}